\documentclass[a4paper,english,aps,manuscript,showpacs]{revtex4}
\usepackage[T1]{fontenc}
\usepackage[utf8]{inputenc}
\usepackage{babel}
\usepackage{amsmath}
\usepackage{amssymb}
\usepackage{graphicx}
\usepackage{esint}
\usepackage[unicode=true,
 bookmarks=true,bookmarksnumbered=false,bookmarksopen=false,
 breaklinks=false,pdfborder={0 0 1},backref=false,colorlinks=false]
 {hyperref}
\hypersetup{pdftitle={Dipole-dipole interactions in nanoplasmonic systems},
 pdfauthor={Marek Nečada, Jani-Petri Martikainen, Päivi Törmä}}
\usepackage{breakurl}



\begin{document}
\newcommand{\ket}[1]{\left|#1\right\rangle }
\newcommand{\bra}[1]{\left\langle #1\right|}
\newcommand{\vect}[1]{\mathbf{\boldsymbol{#1}}}
\newcommand{\ud}{\mathrm{d}}
\newcommand{\modef}{\omega}
\newcommand{\molecf}{\epsilon}
\newcommand{\dipdipint}{g}
\newcommand{\dipmodint}{V}
\newcommand{\perm}{\varepsilon}
\newcommand{\permv}{\perm_{0}}

\title{Quantum emitter dipole-dipole interactions in nanoplasmonic systems}

\author{Marek Ne\v cada}

\author{Jani-Petri Martikainen}

\author{Päivi T\"orm\"a}

\email{paivi.torma@aalto.fi}

\affiliation{COMP Centre of Excellence, Department of Applied Physics, Aalto University,
P.O. Box 15100, Fi-00076 Aalto, Finland}

\date{\today}
\begin{abstract}
We introduce a generalized Dicke-like model to describe two-level
systems coupled with a single bosonic mode. In addition, the two-level
systems mutually interact via direct dipole-dipole interaction. We
apply the model to an ensemble of dye molecules coupled to a plasmonic
excitation in a metallic nanoparticle and study how the dipole-dipole
interaction and configurational randomness introduced to the system
affect the energy spectra. Comparing the system eigenenergies obtained
by our model with the light spectra from a multiple-scattering simulation,
we suggest a way to identify dark modes in our model. Finally, we
perform a parameter sweep in order to determine the scaling properties
of the system and to classify the regions of the parameter space where
the dipole-dipole interactions can have significant effects.
\end{abstract}


\maketitle

Surface plasmon polaritons (SPPs) are hybrid modes of electron plasma
oscillations inside metals and the electromagnetic field inside and
outside of the metallic structure. At resonant frequencies, which
are determined by the plasma frequency of the bulk metal and the nanostructure
geometry, the intensity of the SPP modes in the near field is significantly
magnified. This field enhancement can lead to strong coupling between
the SPP and quantum emitters (QEs) located in the vicinity of the
metal surface, creating a hybrid mode consisting of the SPP and the
QE excitations. The strong coupling manifests itself as an avoided
crossing between the original SPP and QE energy levels, as has been
experimentally realized in various nanoplasmonic structures; see \citep{torma_strong_2015}
and for works that appeared after this review, for instance \citep{chikkaraddy_single-molecule_2016,santhosh_vacuum_2016,zengin_realizing_2015},
and references therein.

Here we address the question whether the dipole-dipole interactions
between the QEs can play a role in such systems. To our knowledge,
there has not been any experimental evidence of this, but theoretical
works exist that admit such possibility.  A study by Salomon \textit{et
al.} \citep{salomon_strong_2012} using the FDTD solution of Maxwell-Liouville
equations for a silver slit array covered by a thin layer of molecules
presented a possibility for an additional mode in the transmission
spectra between the avoided strong coupling modes, provided that either
molecule concentration or their transition dipole moment is large
enough. A multiple-scattering method based on macroscopic quantum
electrodynamics was used by Delga \textit{et al.} \citep{delga_quantum_2014,delga_theory_2014}
to show similar results for a system of single spherical nanoparticle
with adjacent fluorescent molecules. Here we answer this question
by a different approach, namely a modified Dicke model.

Interactions between electromagnetic fields and a collection of QEs
such as atoms or fluorescent molecules are of interest in many areas
of physics. The simplest quantum model of such a system is the original
Dicke model \citep{dicke_coherence_1954}, where identical two-level
systems are coupled to a single harmonic oscillator-like field mode,
all with the same coupling strength, and without direct mutual coupling.
Dicke model can be solved exactly using the algebraic Bethe ansatz
\citep{gaudin_diagonalisation_1976}. There also exists an exact solution
for an extended Dicke model which includes direct coupling term between
the QEs \citep{pan_exact_2005}, but the coupling strengths between
each pair of QEs must be all equal, as well as the QE–field mode couplings,
which is unrealistic for real systems with many QEs. In this article,
we study a model in which all the coupling strengths of both types
of couplings can vary, which is expected to happen in real systems
where the QEs can have various positions and orientations. The price
of relaxing these constraints is the impossibility to diagonalize
the Hamiltonian with the Bethe ansatz, so we use the numerical exact
diagonalization instead.

The original Dicke model describes well a system of atoms in a high-Q
optical nanocavity, as long as the atoms are separated well enough
so that their mutual dipole-dipole interactions are negligible and
the cavity supports single radiation mode near the resonant frequency
of the atoms, and the other modes are well separated. However, our
motivation stems from the study of nanoplasmonic systems where QEs
interact with surface plasmon polariton (SPP) modes supported by a
metallic nanostructure. In the nanoplasmonic systems, the coupling
strengths between the QEs and the SPP vary considerably depending
on the configuration of the QEs.  In particular, we are interested
whether the dipole-dipole couplings between the QEs can have significant
effect on the system.

Approaches different from ours have been developed to model the behaviour
of the nanoplasmonic systems in question. Among the notable ones are
the methods based on finite-difference time-domain (FDTD) solution
of Maxwell-Liuville equations \citep{sukharev_numerical_2011} and
quantum multiple-scattering methods based on macroscopic quantum electrodynamics
\citep{wubs_multiple-scattering_2004}, which have their advantages
and disadvantages. Most notably, they account for the ohmic losses
inside the metal and the consequent line broadening which are significant
in the plasmonic systems. On the other hand, our modified Dicke model
is much less computationally expensive and gives intuitive understanding
of the problem. We compare our model with the multiple-scattering
approach in the Section \ref{sec:Comparison} of this article.

This paper is organized as follows. In Section \ref{sec:The-relaxed-Dicke}
we generalize the extended Dicke model \citep{pan_exact_2005} by
relaxing its equal-coupling symmetries. In Section \ref{sec:The-scattering-approach}
we briefly sketch the principles of the multiple-scattering method,
which is used as a benchmark for our model in Section \ref{sec:Comparison},
where for some example configurations, we compare the resulting energy
levels with the far-field light spectra obtained by multiple-scattering
method as shown by Delga \textit{et al.} \citep{delga_quantum_2014},
observing a clear correspondence between them. Although our model
does not by itself include any information about the visibility of
its eigenenergies, we find an observable which identifies the dark
modes. Finally, in Section \ref{sec:Exact-diagonalization-results}
we use our relaxed Dicke model to perform a parameter sweep in the
single excitation subspace with the goal of identifying the effects
of varying dipole-dipole interactions in the model and their relevance
for the parameters typical in the experiments with quantum emitters
near plasmonic nanostructures. In Section \ref{sec:Conclusions},
we discuss what conditions would the system have to satisfy in order
to make the effects of the dipole-dipole interactions observable.

\section{The relaxed Dicke model\label{sec:The-relaxed-Dicke}}

We consider an ensemble of $K$ identical quantum emitters—modeled
as two-level systems (TLS)—interacting with a single field mode. Our
model Hamiltonian of the system (utilizing the rotating wave approximation)
is 
\begin{eqnarray}
H & = & \hbar\modef\hat{b}^{\dagger}\hat{b}+\sum_{i}\hbar\molecf\left(\hat{S}_{i}^{z}+\frac{1}{2}\right)+\sum_{i}\dipmodint_{i}\left(\hat{b}^{\dagger}\hat{S}_{i}^{-}+\hat{S}_{i}^{+}\hat{b}\right)\nonumber \\
 &  & +\sum_{i<j}\dipdipint_{ij}\left(\hat{S}_{i}^{+}\hat{S}_{j}^{-}+\hat{S}_{j}^{+}S_{i}^{-}\right),\label{eq:Hamiltonian}
\end{eqnarray}
where $\omega$ is the frequency of the field mode, $\epsilon$ is
the resonant frequency of the atoms, $\dipmodint_{i}$ is the coupling
coefficient between the $i$-th atom and the field mode, and $\dipdipint_{ij}$
are the coefficients of the direct interaction between the $i$th
and the $j$th atom. Here $\hat{S}_{i}^{+},\hat{S}_{i}^{-}$ and $\hat{S}_{i}^{z}$
are spin-1/2 operators given by
\begin{eqnarray*}
\hat{S}_{i}^{+} & = & \ket{g_{i}}\bra{e_{i}},\quad\hat{S}_{i}^{-}=\ket{e_{i}}\bra{g_{i}},\\
\hat{S}_{i}^{z} & = & \frac{1}{2}\left(\ket{e_{i}}\bra{e_{i}}-\ket{g_{i}}\bra{g_{i}}\right),
\end{eqnarray*}
where $\ket{g_{i}}$ and $\ket{e_{i}}$ are the ground and excited
states of the $i$th atom; $\hat{b}^{\dagger},\hat{b}$ are the creation
and annihilation operators of the field mode, satisfying the usual
bosonic commutation relation $\left[\hat{b},\hat{b}^{\dagger}\right]=1$.

The extended Dicke model solved by Pan \textit{et al.} \citep{pan_exact_2005}
has the same form as (\ref{eq:Hamiltonian}) but it assumes that the
dipole-dipole and dipole-field couplings are constant, $g_{ij}=g$,
$V_{i}=V$ (in addition, $g_{ij}=0$ in the original Dicke model).
In contrast, in our model the coupling constants in the atom-atom
and atom-field interactions may vary, relaxing the symmetries which
enable the Dicke model to be exactly solvable by the Bethe ansatz.
On the other hand, our model provides much more realistic description
of the physical systems in which we are interested. A typical realization
of our model would consist of quantum emitters (e.g. dye molecules
embedded into polymer matrix) in a cavity or in the vicinity of a
waveguide or a nanoparticle supporting a single dominant EM field
mode—meaning that for at least a certain time scale, the remainder
of the electromagnetic spectrum as well as the non-radiative losses
can be neglected. The quantum emitters are deposited randomly both
in positions and dipole orientations, which leads to some random distribution
of the dipole-dipole coupling strengths\textbf{ $\dipdipint_{ij}$}.

Due to the rotating wave approximation, the total excitation number
operator $\hat{N}=\hat{b}^{\dagger}\hat{b}+\sum_{i}\left(\hat{S}_{i}^{z}+\frac{1}{2}\right)$
commutes with the Hamiltonian. The total excitation number thus remains
conserved and the Hamiltonian can be diagonalized for each excitation
number subspace separately, which allows us to reduce the computational
requirements of diagonalization. In the following, we deal mainly
with the single excitation subspace ($N=1$) which is generated by
the states $\hat{b}^{\dagger}\ket g,\hat{S}_{i}^{+}\ket g$ where
$\ket g=\left(\ket 0\otimes\prod_{j}\ket{g_{j}}\right)$ is the ground
state of the whole system and $\ket 0$ is the vacuum of the bosonic
part. 

Let us describe how the Hamiltonian (\ref{eq:Hamiltonian}) and its
parameters can be derived for a subwavelength-sized system of a plasmonic
resonator and adjacent emitters. The resonator mode alone is obtained
by solving macroscopic Maxwell's equations with the respective constitutive
relations. If the system is not limited to a finite volume, this usually
yields a continuum of mutually orthogonal solutions with infinite
mode volumes (and thus energies). This can be worked around \citep{fredkin_resonant_2003}
by using e.g. a quasistatic approximation (i.e. assuming infinite
speed of light) where the problem is reduced to Gauß law 
\begin{equation}
\nabla\cdot\left(\varepsilon(\omega,\vect r)\vect E(\vect r)\right)=0\label{eq:gauss}
\end{equation}
 with appropriate boundary conditions.  This approximation eliminates
the losses due to the radiation and leads to discrete spectrum in
$\omega$. In order to make the spectrum real, we further neglect
the imaginary part of $\varepsilon(\omega,\vect r)$ for real $\omega$,
thus eliminating the internal losses of the material and making the
plasmonic resonator a closed system. At this point, we get from (\ref{eq:gauss})
a discrete set of quasistatic modes with the dynamics of a harmonic
oscillator, which—after the usual quantization procedure and dropping
the zero point energy—yields the resonator part of the Hamiltonian,
\[
H_{\mathrm{res}}=\hbar\sum_{\lambda}\modef_{\lambda}\hat{b}_{\lambda}^{\dagger}\hat{b}_{\lambda}
\]
and the electric field operator has the form 
\begin{equation}
\hat{\vect E}(\vect r)=\sum_{\lambda}\sqrt{\frac{\hbar\modef_{\lambda}}{2U_{\lambda}\permv}}\left(\vect E_{\lambda}(\vect r)\hat{b}_{\lambda}+\vect E_{\lambda}^{*}\left(\vect r\right)\hat{b}_{\lambda}^{\dagger}\right)\label{eq:Electric operator}
\end{equation}
 where $\omega_{\lambda},\vect E_{\lambda}(\vect r)$ are the solutions
of the classical equation (\ref{eq:gauss}). The operator is normalized
by the quasistatic mode energies $U_{\lambda}=\varepsilon_{0}\int\ud^{3}\vect r\,\left|\vect E_{\lambda}(\vect r)\right|^{2}/2$. 

As for the quantum emitters, we assume they are characterized by two
parameters—the resonant frequency $\epsilon$ and magnitude $\left|\vect{\mu}_{i}\right|$
of their transition dipole moment—and that they have fixed positions
$\vect R_{i}$ and directions of their dipoles $\vect{\mu}_{i}/\left|\vect{\mu}_{i}\right|$.
Hamiltonian for the QEs before introducing interactions is $H_{\mathrm{QE}}=\sum_{i}\hbar\molecf\left(\hat{S}_{i}^{z}+\frac{1}{2}\right)$
and its dipole moment operator 
\begin{equation}
\hat{\vect{\mu}}_{i}=\vect{\mu}_{i}\left(\hat{S}_{i}^{+}+\hat{S}_{i}^{-}\right).\label{eq:dipole moment operator}
\end{equation}
 Their dipoles interact with the resonator's electric field via the
term 
\begin{equation}
H_{\mathrm{res\text{–QE}}}=-\sum_{i}\hat{\vect{\mu}}_{i}\cdot\hat{\vect E}(\vect R_{i})\label{eq:NP QE interaction}
\end{equation}
 and with each other via the quasistatic dipole-dipole interaction
\[
H_{\mathrm{QE\text{–}QE}}=\frac{1}{4\pi\varepsilon_{0}}\sum_{i<j}\left(\frac{\hat{\vect{\mu}}_{i}\cdot\hat{\vect{\mu}}_{j}}{\left|\vect R_{i}-\vect R_{j}\right|^{3}}-3\frac{\left[\hat{\vect{\mu}}_{i}\cdot\left(\vect R_{i}-\vect R_{j}\right)\right]\left[\hat{\vect{\mu}}_{j}\cdot\left(\vect R_{i}-\vect R_{j}\right)\right]}{\left|\vect R_{i}-\vect R_{j}\right|^{5}}\right).
\]
Finally, we take $H=H_{\mathrm{res}}+H_{\mathrm{QE}}+H_{\mathrm{QE\text{–QE}}}+H_{\mathrm{res\text{–}QE}}$
and perform the rotating wave approximation, dropping all the terms
containing $\hat{b}_{\lambda}\hat{S}_{i}^{-},\hat{b}_{\lambda}^{\dagger}\hat{S}_{i}^{+},\hat{S}_{i}^{-}\hat{S}_{j}^{-}$
or $\hat{S}_{i}^{+}\hat{S}_{j}^{+}$, obtaining 
\begin{eqnarray}
H & = & \sum_{\lambda}\hbar\modef_{\lambda}\hat{b}_{\lambda}^{\dagger}\hat{b}_{\lambda}+\sum_{i}\hbar\molecf\left(\hat{S}_{i}^{z}+\frac{1}{2}\right)+\sum_{i,\lambda}\dipmodint_{i\lambda}\left(\hat{b}_{\lambda}^{\dagger}\hat{S}_{i}^{-}+\hat{S}_{i}^{+}\hat{b}_{\lambda}\right)\nonumber \\
 &  & +\sum_{i<j}\dipdipint_{ij}\left(\hat{S}_{i}^{+}\hat{S}_{j}^{-}+\hat{S}_{j}^{+}\hat{S}_{i}^{-}\right),\label{eq:Hamiltonian-1}
\end{eqnarray}
 with the coupling coefficients 
\[
\dipmodint_{i\lambda}=\sqrt{\frac{\hbar\omega_{\lambda}}{2U_{\lambda}\varepsilon_{0}}}\hat{\vect{\mu}}_{i}\cdot\hat{\vect E}_{\lambda}(\vect R_{i})
\]
and

\[
\dipdipint_{ij}=\frac{1}{4\pi\varepsilon_{0}}\left(\frac{\hat{\vect{\mu}}_{i}\cdot\hat{\vect{\mu}}_{j}}{\left|\vect R_{i}-\vect R_{j}\right|^{3}}-3\frac{\left[\hat{\vect{\mu}}_{i}\cdot\left(\vect R_{i}-\vect R_{j}\right)\right]\left[\hat{\vect{\mu}}_{j}\cdot\left(\vect R_{i}-\vect R_{j}\right)\right]}{\left|\vect R_{i}-\vect R_{j}\right|^{5}}\right).
\]

For practical calculations, it is usually not necessary to take into
account all the resonator modes: only the modes with frequencies near
enough to the resonant frequency of the QEs will significantly affect
the system. Assuming there is only one such significant mode (which
is also the assumption of the Dicke model), we arrive at the Hamiltonian
(\ref{eq:Hamiltonian}).

\section{The scattering approach\label{sec:The-scattering-approach}}

In the following, we  compare the results of our model explained above
to a multiple scattering model described in \citep{wubs_multiple-scattering_2004,delga_quantum_2014}.
For details, we refer the reader to the supplement of \citep{delga_quantum_2014},
but we outline the main properties of the model here. 

The approach is based on macroscopic quantum electrodynamics, where
medium (including the plasmonic resonator in our case) is modelled
by a continuum of harmonic oscillators coupled to the microscopic
elecromagnetic fields in a manner that (before quantization) reproduces
the phenomenological constitutive relations; the electromagnetic field
operators can be then expressed in terms of the classical dyadic Green's
functions \citep{gruner_green-function_1996}. Adding the microscopic
QEs and coupling them to these quantized fields gives rise to a Lippmann-Schwinger
equation which is hard to solve if the QEs are two-level systems.
Therefore, the two-level systems are approximated by harmonic oscillators,
at the cost of reliability  of the model for problems that involve
multiple excitations. For a given initial state, e.g. some of the
emitters excited, we can derive a light spectrum that can be detected
in an arbitrary point of space. 

In contrast to our model described in the previous  section, this
approach has some advantages: it gives the light spectrum—a quantity
of direct experimental relevance, it accounts for the absorption in
the media, and it includes the field retardation effects, keeping
its validity at longer-than-wavelength scales. These features make
it a good benchmark for our model. On the other hand, the multiple-scattering
approach will fail to describe the effects of higher level of excitation
and it is significantly slower computationally, as it requires solving
a separate matrix inversion problem for each sample frequency of the
outgoing light spectrum, each inversion having the same time complexity
as one exact diagonalization.

\section{Comparison\label{sec:Comparison}}

\begin{figure}
\includegraphics[width=0.3\columnwidth]{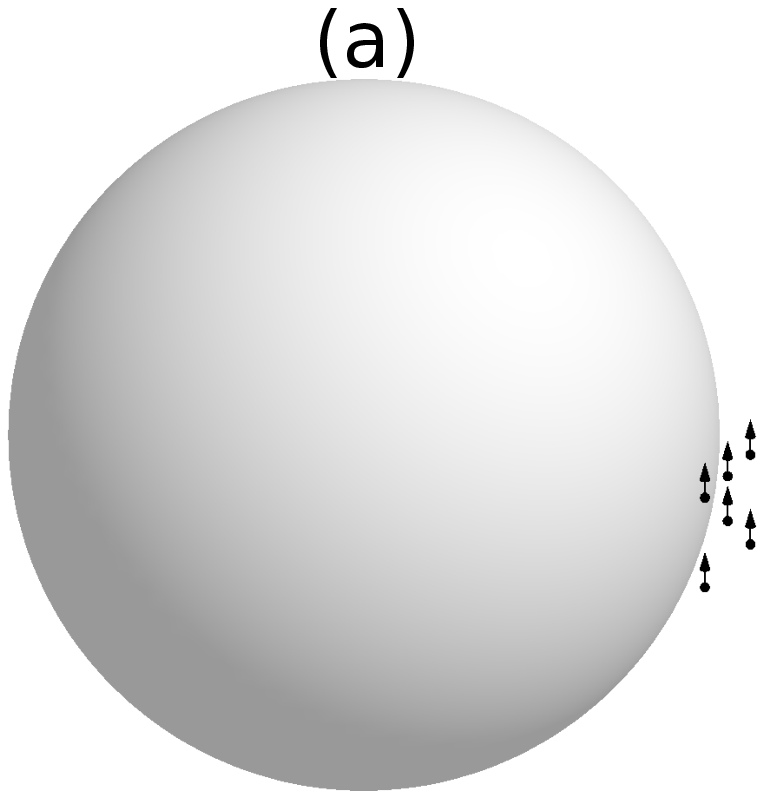} 
\includegraphics[height=5cm]{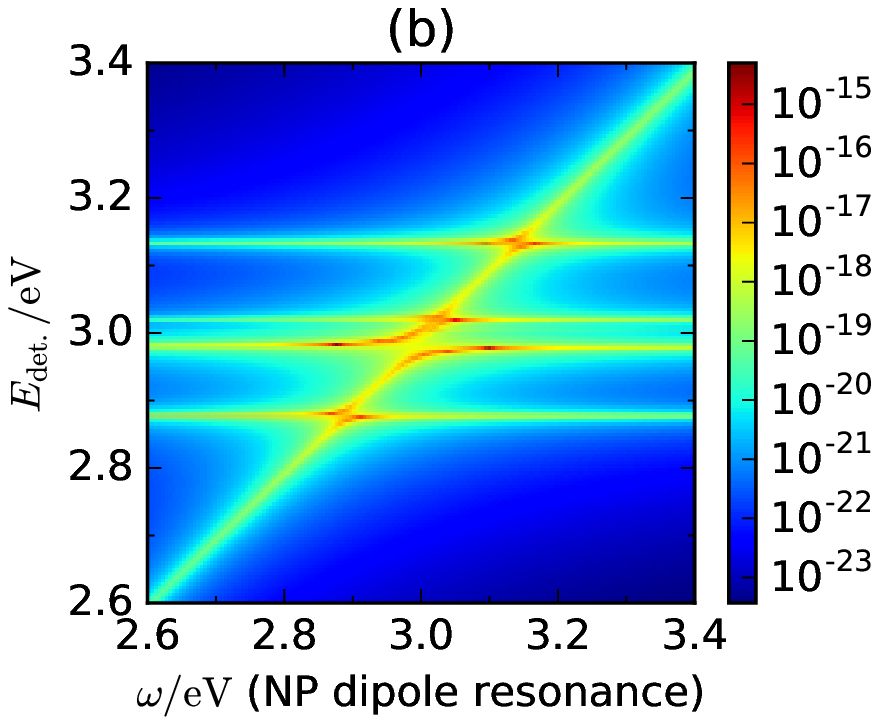} 
\includegraphics[height=5cm]{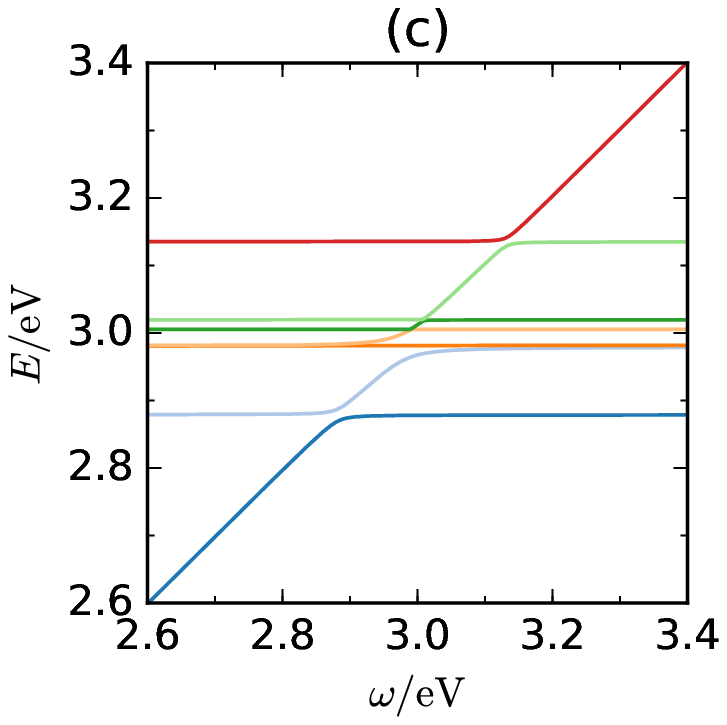}
\caption{Configuration of the system (a) and the corresponding spectra, varying
the plasma frequency of the metal: (b) far-field light spectrum obtained
by the multipole scattering method, (c) eigenvalues of the Hamiltonian
(\ref{eq:Hamiltonian}). \label{fig:comparison spectra}}
\end{figure}

Let us take a simple physical system in order to compare our modified
Dicke model and the multiple scattering approach. The system consists
of a metal nanosphere, and several molecular dipoles modelled as two-level
systems nearby, as in Figure \ref{fig:comparison spectra}. The relative
permittivity of the nanosphere is approximated by the Drude model,
$\perm(\omega)=\perm_{\infty}-\frac{\omega_{p}^{2}}{\omega\left(\omega+i\gamma_{p}\right)}$
with parameters  $\perm_{\infty}=4.6$, $\hbar\gamma_{p}=0.001\,\mathrm{eV}$
(the plasma frequency is chosen arbitrarily low in order to create
peaks comparable to the modified Dicke model spectrum), its radius
is $r=7\,\mathrm{nm}$, and we vary its plasma frequency $\omega_{p}$.
In the quasistatic approximation \citep{ruppin_decay_1982}, $l$-th
order electric multipole resonances of a nanosphere are determined
by the equation $0=\perm(\omega_{l})+\perm_{\mathrm{b}}(l+1)/l$
where $\perm_{\mathrm{b}}$ is the environment relative permittivity.
In our case, the dipole resonance is thus located at $\omega_{1}=3.02\,\mathrm{eV/\hbar}$—we
use this value as the mode frequency $\modef$ in the Hamiltonian
(\ref{eq:Hamiltonian}), whereas for the multipole scattering method,
we use Mie theory with the aforementioned parameters to calculate
the nanosphere response. 

The molecules have transition dipole moments of $\left|\vect{\mu}\right|=0.19\,\mathrm{eV\cdot nm}$,
they are aligned in the $z$-direction, and positioned in a plane
$8\,\mathrm{nm}$ from the centre of the nanosphere. These values
were chosen in order to make the molecular interactions significant.
For simplicity, we include only the electric dipole response of the
nanosphere, neglecting all the higher multipole terms of Mie theory,
and we assume that the field has the same shape as it would have in
the electrostatic case of a polarized sphere, i.e. we neglect the
outward radiation. Moreover, here we place the molecules near the
equatorial plane perpendicular to the $z$ axis in order to keep the
interactions the $x$ and $y$ dipoles of the nanoparticle negligible.
Therefore, we can model the system with the Hamiltonian (\ref{eq:Hamiltonian})
(otherwise the more general Hamiltonian (\ref{eq:Hamiltonian-1})
would be needed). 

In Figure \ref{fig:comparison spectra} we show the light spectrum
obtained by the multiple scattering method at a point located  $10\,\mathrm{\mu m}$
away from the centre of the nanoparticle together with the eigenenergies
obtained from our model. There is a clear correspondence between the
peaks of the light spectrum and the eigenenergies from the respective
approaches. However, not every energy level has its corresponding
peak in the light spectrum; we discuss these dark states in the next
section.

\subsection*{Characterization of the dark modes}

Hamiltonian (\ref{eq:Hamiltonian}) describes a closed quantum system
of electric dipoles with Coulombic interaction and certain internal
dynamics, and by itself does not carry any information about interaction
with radiation, hence nor about the visibility of its eigenstates.
Therefore, we extend the system to include radiation modes and assess
their visibility using perturbation theory. Let the new Hamiltonian
be
\begin{equation}
H'\equiv H_{0}+V=H+\sum_{\underline{\vect k}}\hbar\omega_{\vect k}a_{\underline{\vect k}}^{\dagger}a_{\underline{\vect k}}+V_{\mathrm{ext}}\label{eq:radiation hamiltonian}
\end{equation}
where $\underline{\vect k}\equiv(\vect k,\iota)$ labels the transversal
(radiation) modes with mutually orthogonal wave and polarization vectors
$\vect k$ and $\vect{\varepsilon}_{\vect k,\iota}$ (here $\iota=1,2$
labels the polarisation basis vectors), $\omega_{\vect k}=c\left|\vect k\right|$
is the corresponding mode frequency, and $V_{\mathrm{ext}}$ is the
interaction between the transversal modes and all the dipoles (including
the nanoparticle)
\begin{eqnarray}
V_{\mathrm{ext}} & =- & \hat{\vect E}^{\bot}(\vect r=0)\cdot\hat{\vect{\mu}}_{\mathrm{tot.}}\label{eq:radiation interaction}\\
 & = & \sum_{\underline{\vect k}}i\sqrt{\frac{\hbar\omega_{\vect k}}{2\permv L^{3}}}\left(\vect{\varepsilon}_{\underline{\vect k}}a_{\underline{\vect k}}-\vect{\varepsilon}_{\underline{\vect k}}a_{\underline{\vect k}}^{\dagger}\right)\cdot\left(\vect{\mu}_{\mathrm{NP}}\left(b+b^{\dagger}\right)+\sum_{i}\vect{\mu}_{i}\left(S_{i}^{+}+S_{i}^{-}\right)\right).
\end{eqnarray}
Here we use the usual way to quantize transversal EM modes \citep{cohen-tannoudji_processus_2000}.
 Furthemore, we assume a cutoff in the frequencies such that $\vect k\cdot\vect r_{\mathrm{sys}}\ll1$
where $\vect r_{\mathrm{sys}}$ is the radius of the volume in which
the dipoles are placed, so all the dipole positions can be replaced
with $\vect r=0$ (dipole approximation for the whole original system)
and ultraviolet divergences are avoided. Finally, we assume that all
the conditions to apply the Fermi's golden rule are fulfilled. Decay
rate of an initial state $\ket{\alpha}$ from the eigenspace of the
original Hamiltonian $H$ into the continuum of final transversal
photonic states $\ket f$ is proportional to the sum of squares of
the corresponding transition amplitudes \citep{cohen-tannoudji_processus_2000}
\begin{equation}
\Gamma_{\alpha}\propto\sum_{f}\left|\bra fV_{\mathrm{ext}}\ket{\alpha}\right|^{2}.\label{eq:decay rate prop}
\end{equation}

To keep the length of the formulae reasonable, in the following we
denote $\vect{\mu}_{0}\equiv\vect{\mu}_{\mathrm{NP}}$, $S_{0}^{+}\equiv b^{\dagger}$,
$S_{0}^{-}\equiv b$. In the single excitation subspace (using RWA),
both initial and final subspaces are spanned by the states obtained
by applying a single corresponding creation operator onto the vacuum
state of $H_{0}$, $\ket f=\ket{\underline{\vect k}}\equiv a_{\underline{\vect k}}^{\dagger}\ket 0$,
$\ket{\alpha}=\sum_{i=0}^{K}c_{i}S_{i}^{+}\ket 0$. Substituting this
to (\ref{eq:decay rate prop}) gives 
\begin{eqnarray*}
\Gamma_{\alpha} & \propto & \sum_{\underline{\vect k}}\bra{\alpha}\hat{\vect{\mu}}_{\mathrm{tot.}}\cdot\sum_{\underline{\vect k'}}i\sqrt{\frac{\hbar\omega_{\vect k'}}{2\permv L^{3}}}\left(\vect{\varepsilon}_{\underline{\vect k'}}a_{\underline{\vect k'}}-\vect{\varepsilon}_{\underline{\vect k'}}a_{\underline{\vect k'}}^{\dagger}\right)a_{\underline{\vect k}}^{\dagger}\\
 &  & \times\ket 0\bra 0a_{\underline{\vect k}}\sum_{\underline{\vect k''}}-i\sqrt{\frac{\hbar\omega_{\vect k''}}{2\permv L^{3}}}\left(\vect{\varepsilon}_{\underline{\vect k''}}a_{\underline{\vect k''}}-\vect{\varepsilon}_{\underline{\vect k''}}a_{\underline{\vect k''}}^{\dagger}\right)\cdot\hat{\vect{\mu}}_{\mathrm{tot.}}\ket{\alpha}\\
 & = & -\sum_{\underline{\vect k}}\frac{\hbar\omega_{\vect k}}{2\permv L^{3}}\bra{\alpha}\hat{\vect{\mu}}_{\mathrm{tot.}}\cdot\vect{\varepsilon}_{\underline{\vect k}}a_{\underline{\vect k}}a_{\underline{\vect k}}^{\dagger}\ket 0\bra 0a_{\underline{\vect k}}a_{\underline{\vect k}}^{\dagger}\vect{\varepsilon}_{\underline{\vect k}}\cdot\hat{\vect{\mu}}_{\mathrm{tot.}}\ket{\alpha}
\end{eqnarray*}
Here we used the commutativity of the photonic operators $a_{\underline{\vect k}},a_{\underline{\vect k}}^{\dagger}$
with the $S_{i}^{\pm}$ operators contained in $\hat{\vect{\mu}}_{\mathrm{tot.}}$
together with the fact that $a_{\underline{\vect k}}a_{\underline{\vect k''}}\ket{\alpha}=0$
and $\bra 0a_{\underline{\vect k}}a_{\underline{\vect k''}}^{\dagger}\ket 0$
is nonzero only if $\underline{\vect k''}=\underline{\vect k}$. We
assume that the space supporting the radiation modes is spherically
symmetric, hence for the sum over $\underline{\vect k}$ we get
\begin{flalign*}
\sum_{\underline{\vect k}} & \omega_{\vect k}\vect{\varepsilon}_{\underline{\vect k}}a_{\underline{\vect k}}a_{\underline{\vect k}}^{\dagger}\ket 0\bra 0a_{\underline{\vect k}}a_{\underline{\vect k}}^{\dagger}\vect{\varepsilon}_{\underline{\vect k}}\\
 & =\sum_{\underline{\vect k}}\omega_{\vect k}\vect{\varepsilon}_{\underline{\vect k}}\ket 0\bra 0\vect{\varepsilon}_{\underline{\vect k}}\\
 & =\ket 0\bra 0\sum_{k}\omega_{k}\sum_{\iota=1,2}\int\ud\Omega\,\vect{\varepsilon}_{\vect k,\iota}\vect{\varepsilon}_{\vect k,\iota}\\
 & =\frac{8}{3}\pi\vect I\ket 0\bra 0\sum_{k}\omega_{k},
\end{flalign*}
i.e. a multiple of the unit tensor $\vect I$ (one way to calculate
the angular integral $\int\ud\Omega\,\vect{\varepsilon}_{\vect k,\iota}\vect{\varepsilon}_{\vect k,\iota}$
in the last step is to choose the unit vectors tangential to the circles
of latitude and longitude for the polarisation vectors: $\vect{\varepsilon}_{\vect k,1}=\vect{\hat{\theta}},\vect{\varepsilon}_{\vect k,2}=\vect{\hat{\phi}}$).
Therefore, 
\begin{eqnarray*}
\Gamma_{\alpha} & \propto & \bra 0\left(\sum_{i=0}^{K}c_{i}^{*}S_{i}^{-}\right)\sum_{j=0}^{K}\vect{\mu}_{j}\left(S_{j}^{+}+S_{j}^{-}\right)\ket 0\\
 &  & \cdot\bra 0\sum_{m=0}^{K}\vect{\mu}_{m}\left(S_{m}^{+}+S_{m}^{-}\right)\sum_{n=0}^{K}c_{n}S_{n}^{+}\ket 0\\
 & = & \bra 0\sum_{i=0}^{K}c_{i}^{*}S_{i}^{-}\sum_{j=0}^{K}\vect{\mu}_{j}S_{j}^{+}\ket 0\cdot\bra 0\sum_{m=0}^{K}\vect{\mu}_{m}S_{m}^{-}\sum_{n=0}^{K}c_{n}S_{n}^{+}\ket 0.
\end{eqnarray*}
All the nonzero terms of expressions on the sides of the projector
$\ket 0\bra 0$ are just multiples of the vacuum state, so the projector
can be put away,
\[
\Gamma_{\alpha}\propto\bra{\alpha}\sum_{j=0}^{K}S_{j}^{+}\vect{\mu}_{j}\cdot\sum_{m=0}^{K}S_{m}^{-}\vect{\mu}_{m}\ket{\alpha}\equiv\bra{\alpha}\hat{P}\ket{\alpha}.
\]
The radiative decay rates are thus up to a constant factor given by
the expectation value of an observable $\hat{P}$. The operator 
\begin{equation}
\hat{P}=\sum_{j=0}^{K}S_{j}^{+}\vect{\mu}_{j}\cdot\sum_{m=0}^{K}S_{m}^{-}\vect{\mu}_{m}\label{eq:The Observable-1}
\end{equation}
 resembles the total dipole moment squared, but it is not equal to
the operator $\hat{\vect{\mu}}_{\mathrm{tot.}}^{2}$ which contains
a positive offset caused by the presence of terms like $S_{j}^{-}S_{j}^{+}$,
causing an overall positive shift and therefore its expectation value
being always positive. On the other hand, $\bra{\alpha}\hat{P}\ket{\alpha}$
can be zero, which means that the state $\ket{\alpha}$ does not radiate
(in the given approximation), i.e. it is a dark state.

\begin{figure}
\includegraphics[bb=153bp 282bp 547bp 672bp,clip,width=1\columnwidth]{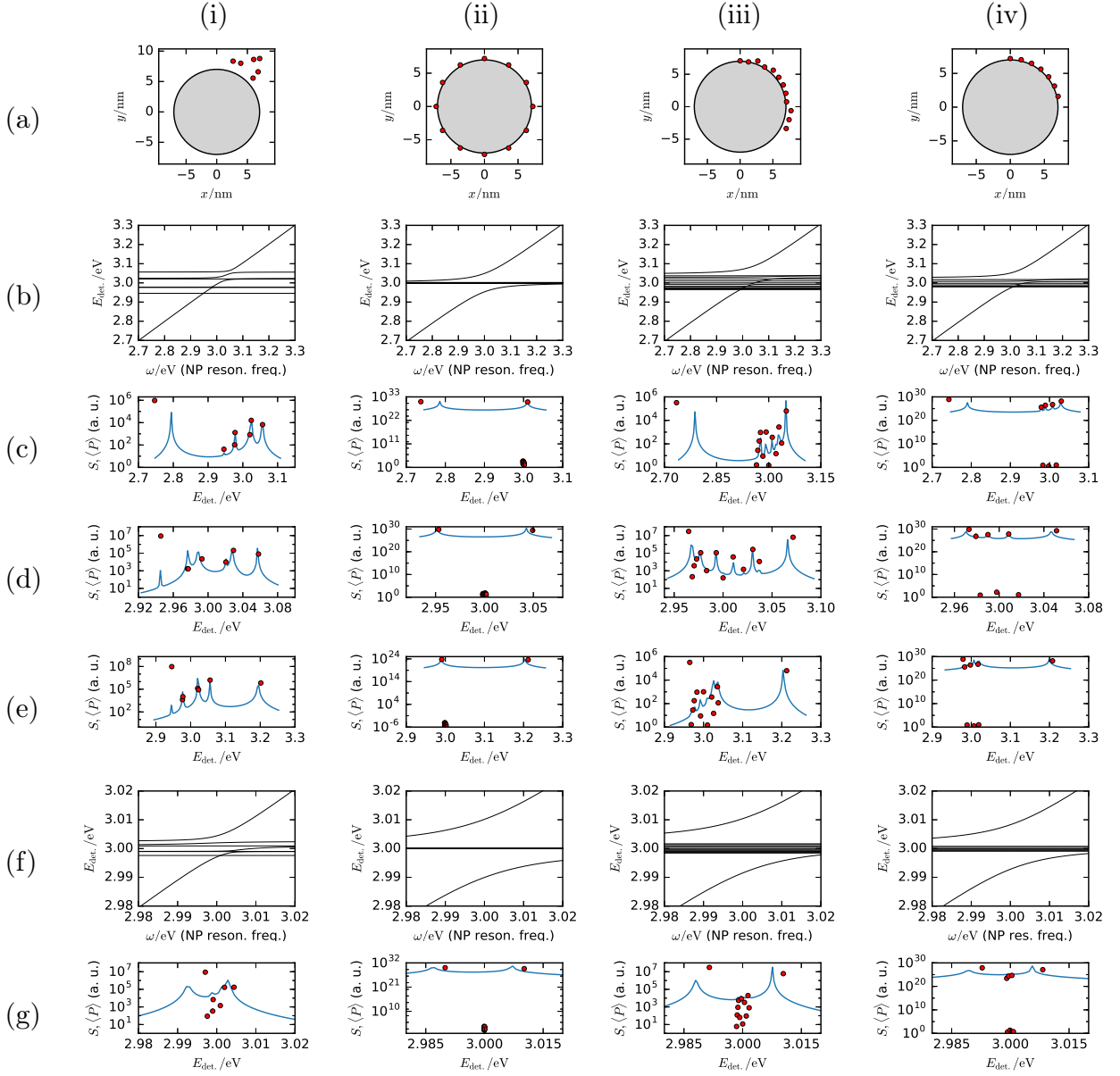}

\caption{Example configurations (a) of quantum emitters in the equatorial plane
of a spherical nanoparticle (with dipoles oriented perpendicular to
the plane), the corresponding energy spectra obtained from the modified
Dicke model for varying nanoparticle resonance (b), and light spectra
from the multiple scattering model (blue lines, arbitrary units) together
with the expectation values of the observable $\hat{P}$ for the energy
eigenstates (red dots, arbitrary units) for the nanoparticle dipole
resonance set at 2.8 eV (c), 3.0 eV (d) and 3.2 eV (e). The single
QEs have transition energy $\hbar\epsilon=3.0\,\mathrm{eV}$ and dipole
moment (c,d,e) $\mu=0.19\,\mathrm{eV\cdot nm=9.1\, D}$; the Drude
damping is set to $\hbar\gamma_{\mathrm{P}}=1\,\mathrm{meV}$. Rows
(f,g) are analogous to rows (b,d) but for a lower dipole moment $\mu=0.04\,\mathrm{eV\cdot nm=1.9\, D}.$
\label{fig:dark modes and observable}}
\end{figure}

Figure \ref{fig:dark modes and observable}(c,d,e) shows the expectation
values $\bra{\alpha}\hat{P}\ket{\alpha}$ for the eigenstates of $H$
in the example configurations. Those states for which the expectation
value is very low are indeed dark also in the results of the multiple
scattering method. Moreover, the relative intensities of the brighter
states correspond well to each other, $S(\alpha)/S(\beta)\approx\bra{\alpha}\hat{P}\ket{\alpha}/\bra{\beta}\hat{P}\ket{\beta}$,
if the compared eigenstates $\ket{\alpha},\ket{\beta}$ are well separated
from others (otherwise their contributions in the total light spectrum
cannot be distinguished) and if they do not contain significant contribution
from the nanoparticle dipole (the emission properties of the two types
of dipoles differ because of the different internal loss channels,
which are however not considered in our model). This is demonstrated
in Fig. \ref{fig:dark modes and observable}(i) where the QEs are
further away from the nanoparticle than in the other examples and
they are all very close to each other, so their mutual dipole-dipole
couplings $g_{ij}$ are stronger than their couplings with the nanoparticle
$V_{i}$.

\section{Exact diagonalization results\label{sec:Exact-diagonalization-results}}

In this section, we vary the parameters of the model Hamiltonian (\ref{eq:Hamiltonian})
in a systematic way and compute the energy spectrum by exact diagonalization.
Again, we focus only on the single-excitation subspace, i.e. the energy
eigenstates satisfying $N=1$. Such states can be expressed in the
form $\ket{\psi}=\left(C\hat{b}^{\dagger}+\sum_{i}c_{i}\hat{S}_{i}^{+}\right)\ket g$
 where $c_{i},C$ are still complex coefficients.

\emph{}

Here we use a different way of determining $V_{i}$ than in the benchmark
calculation above. We assume that the mode field is homogeneous in
the volume of interest (where the QEs are), with the intensity $\vect E$
being one of the parameters.  The coupling constants are again determined
as $V_{i}=-\vect{\mu}_{i}\cdot\vect E$ and are therefore dependent
mainly on the orientation of a given dipole.

\begin{figure*}
\includegraphics[width=0.49\columnwidth]{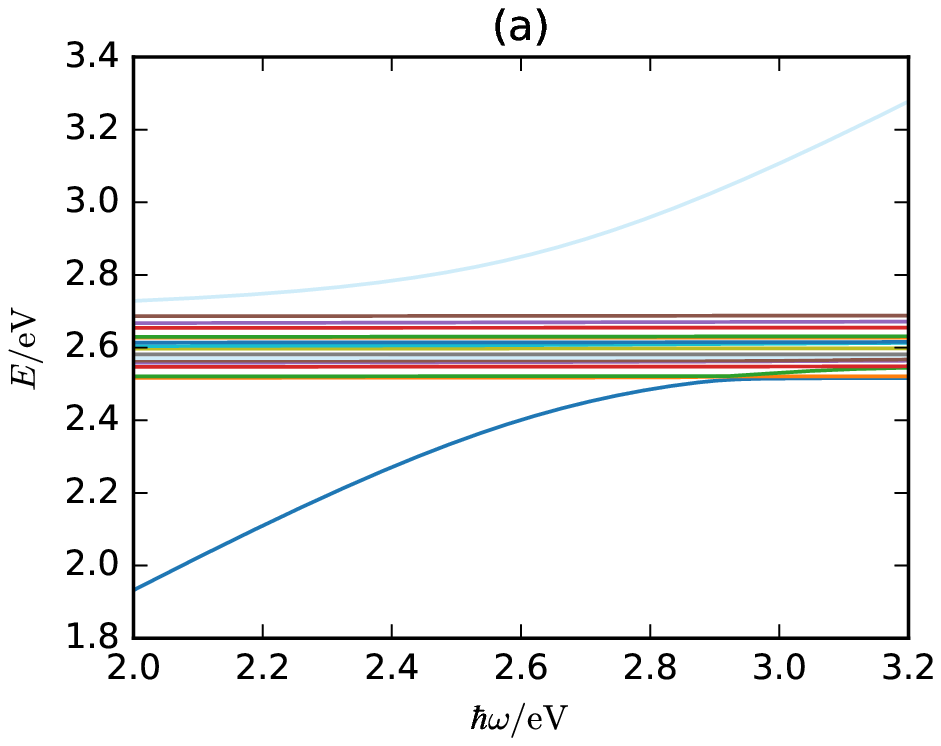} 
\includegraphics[width=0.49\columnwidth]{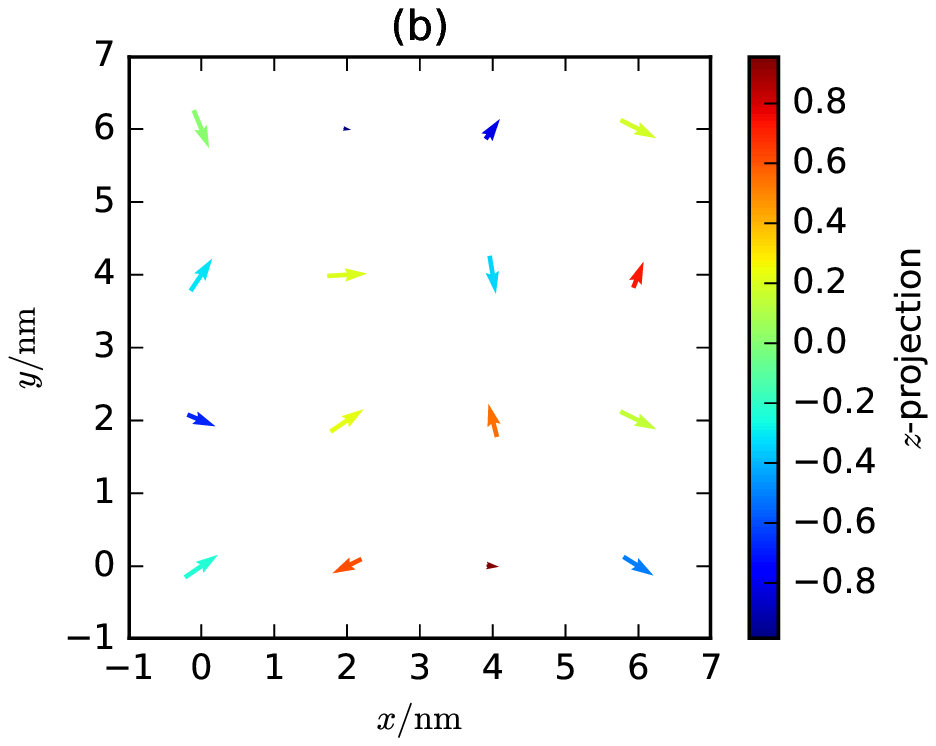}
\caption{Sample configuration of the dipoles (b) and corresponding energy
spectrum of the single-excitation subspace (a) of the modified Dicke
model, with varying energy of the bosonic mode. The parameters used
here are $E=E_{z}=2.4\cdot10^{8}\,\mathrm{V\, m^{-1}}$, $\epsilon=2.6\,\mathrm{eV}/\hbar,\mu=20\,\mathrm{D}$.
\label{fig:Sample configuration and spectrum} Here the dipoles are
located in a regular $4\times4$ square lattice but are randomly oriented.
Two main polariton branches appear together with a central band of
energies around the original transition frequency of a single molecule.
The width of the central band grows with the dipole-dipole couplings
$g_{ij}$.}
\end{figure*}

In the following, we study the energy spectra for varying energy $\omega$
of the bosonic field, keeping the free TLS energy difference $\epsilon$
fixed. This captures the possibility to tune the field mode, whereas
the spectral properties of a molecule are given. A sample spectrum
is shown in Fig. \ref{fig:Sample configuration and spectrum}. Regardless
of the specific configuration, in the single-excitation subspace there
will generally be a bounded ``band'' of $K-2$ eigenvalues (where
$K$ is the number of QEs) nearby the original $\epsilon$, and two
``polariton branches'' below and above the band asymptotically approaching
$\omega$ for $\omega\ll\epsilon$ and $\omega\gg\epsilon$, respectively.
The exact positions of the eigenvalues inside the band depend nontrivially
on the configuration of the dipoles, but there are several quantitative
attributes of the shape of the spectra—e.g. the position and the width
of the band or the separation of the polariton branches—whose dependence
on some basic parameters can be studied statistically. 

Some observations about the spectra follow directly from the structure
of the Hamiltonian (\ref{eq:Hamiltonian}). For small values of $V_{i}$,
the width of the central band in the spectra is directly proportional
to the dipole-dipole couplings $g_{ij}$. Therefore the width scales
with the dipole moment as $\mu^{2}$ and with the length scale $l$
(proportional to the interparticle distances) as $l^{-3}$, and hence
it grows linearly with the concentration of the molecules if the other
parameters stay fixed. When the dipole-dipole couplings $g_{ij}$
are large enough, the lower polariton branch might cross some of the
central band energies for $\omega>\epsilon$, as can be seen in Fig.
\ref{fig:Sample configuration and spectrum}.

The magnitude of the dipole-field couplings $V_{i}$ then affects
mainly the mutual separation of the polariton branches. For small
enough dipole-dipole couplings $g_{ij}$ (so that the band stays well
between the polariton branches) the lower polariton branch approaches
$\epsilon$ for large $\omega$, whereas there is a certain gap between
the upper branch and $\epsilon$ for small $\omega$. We observe that
the computed polariton branches fit quite well onto the formula
\begin{equation}
\tilde{\omega}_{\pm}^{2}=\frac{1}{2}\left[\omega^{2}+\omega_{0}{}^{2}+\Omega^{2}\pm\sqrt{\left(\omega^{2}+\omega_{0}{}^{2}+\Omega^{2}\right)^{2}-4\omega^{2}\omega_{0}^{2}}\right].\label{eq:Hopfield dispersion}
\end{equation}
Such a dependence is found in the dispersion relations derived from
several models of propagating waves (e.g. surface plasmon polaritons
\citep{torma_strong_2015} or usual electromagnetic plane waves \citep{hopfield_theory_1958})
interacting with emitters distributed homogeneously in the direction
of wave propagation and without dipole-dipole interactions. In that
context, $\omega$ from (\ref{eq:Hopfield dispersion}) is the frequency
of the wave of a particular wavelength in the absence of the emitters,
$\omega_{0}$ is the transition frequency of the uncoupled emitters
and $\Omega^{2}$ is a quantity linearly proportional to the polarisability
of the emitters and also to their concentration; $\Omega$ is the
Rabi splitting, equal to the difference between the polariton branches
$\tilde{\omega}_{+}-\tilde{\omega}_{-}$ at resonance ($\omega=\omega_{0}$),
and $\sqrt{\omega_{0}^{2}+\Omega^{2}}$ corresponds to the low-energy
asymptote of the upper polariton branch.

For larger $g_{ij}$, the lower polariton branch starts to cross some
of the band levels and therefore fitting the lowest eigenvalue onto
$\tilde{\omega}_{-}$ is no longer reliable. Nevertheless, using only
the upper branch for the fit yields still reliable results even for
$\omega_{0}$.  As could be seen in Fig. \ref{fig:dark modes and observable}
(i,iii,iv), the shape of the lower polariton branch may still be apparent
in the spectrum of the Hamiltonian even if it penetrates the central
band.

In the following, we will see that neither of the parameters $\omega_{0},\Omega$
depend significantly on the dipole-dipole couplings $g_{ij}$ (cf.
the section Scaling effects and Fig. (\ref{fig:Scaling-mu})). The
polariton splitting $\Omega$ does, however, scale with the single
emitter dipole moment magnitude $\mu$ and the number of dipoles $N$
as $\Omega\propto\sqrt{\mu^{2}N}$. Therefore, the width of the central
band will grow faster than the polariton splitting when the dipole
concentration is increased.

\begin{figure*}
\noindent
\includegraphics[clip,width=0.24\textwidth]{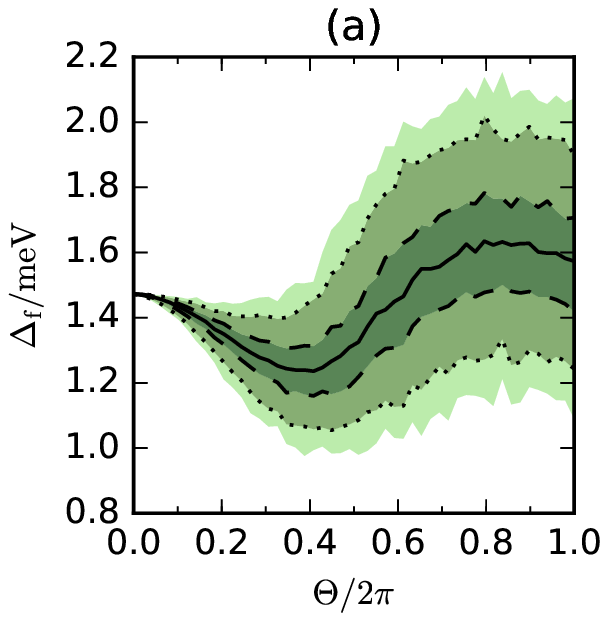}  
\includegraphics[width=0.24\textwidth]{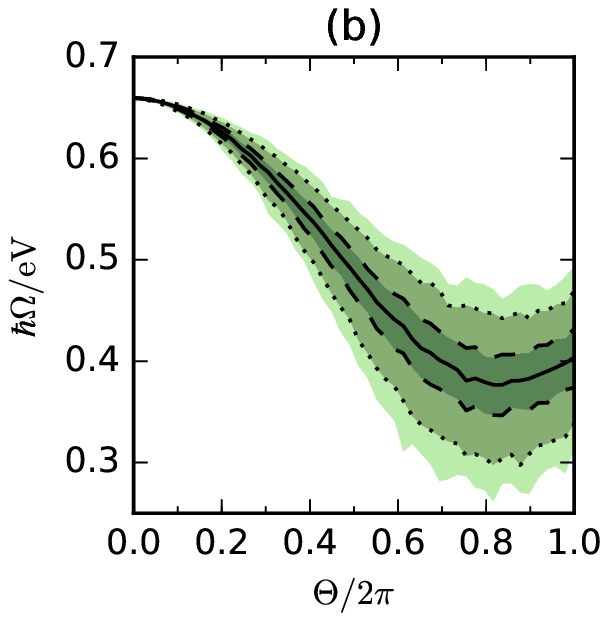}
\\
\includegraphics[width=0.24\textwidth]{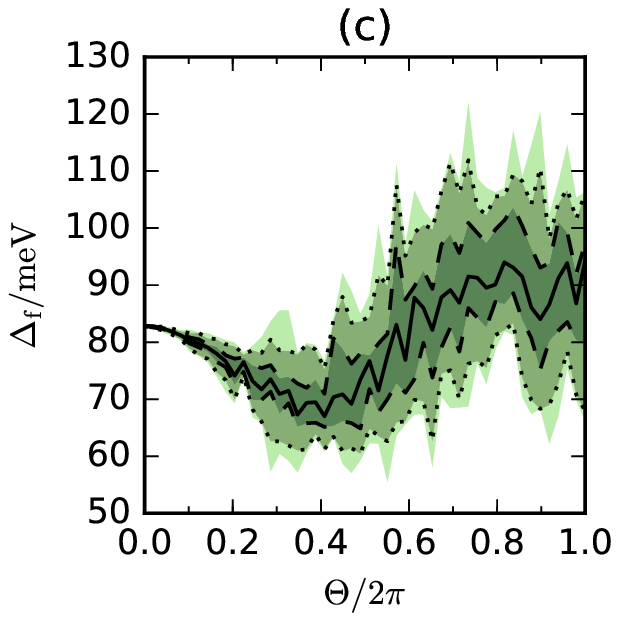}  
\includegraphics[width=0.24\textwidth]{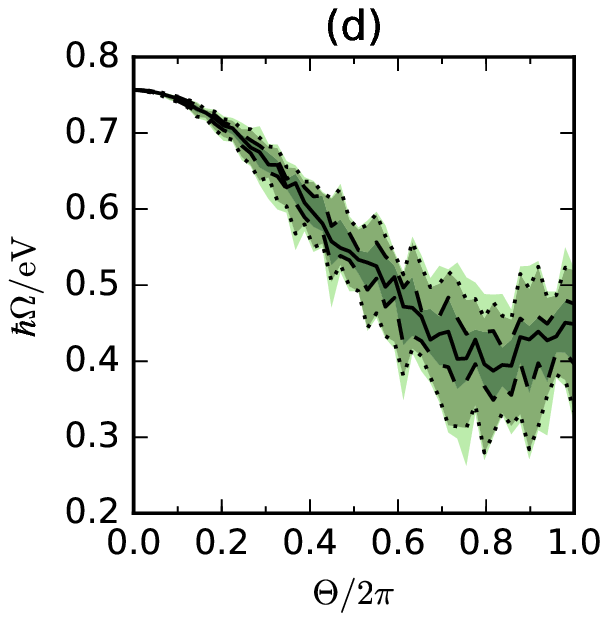}
\caption{Effects of angular randomness on the spectrum in a $4\times4$ dipole
array ($16\,\mathrm{nm\times16\,\mathrm{nm}}$): (a,c) width of the
dipole band $\Delta_{\mathrm{f}}$ taken as the difference between
the second highest and the second lowest eigenvalue at $\omega=\epsilon=2.6\,\mathrm{eV}$;
(b,d) polariton splitting $\Omega$ from fitting the relation (\ref{eq:Hopfield dispersion}).
The parameters are (a,c) $\mu=2\,\mathrm{D}$, $E=E_{z}=2.4\cdot10^{9}\,\mathrm{V\, m^{-1}}$
and (b,d) $\mu=15\,\mathrm{D}$, $E=E_{z}=3.2\cdot10^{8}\,\mathrm{V\, m^{-1}}$
(in both cases $\mu E_{z}=0.1\,\mathrm{eV}$). The lines correspond
to the quantiles 0.01, 0.05, 0.25, 0.5, 0.75, 0.95, 0.99, i.e. the
areas delimited by light, medium and dark shades delimit 98 \%, 90
\% and 50 \% of the values, respectively.\emph{\label{fig:angleRand}}}
\end{figure*}
\begin{figure*}
\noindent
\includegraphics[width=0.3\textwidth]{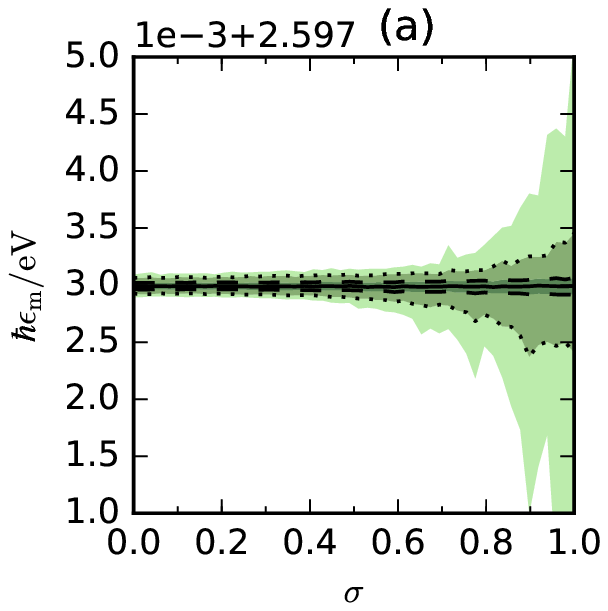} 
\includegraphics[width=0.3\textwidth]{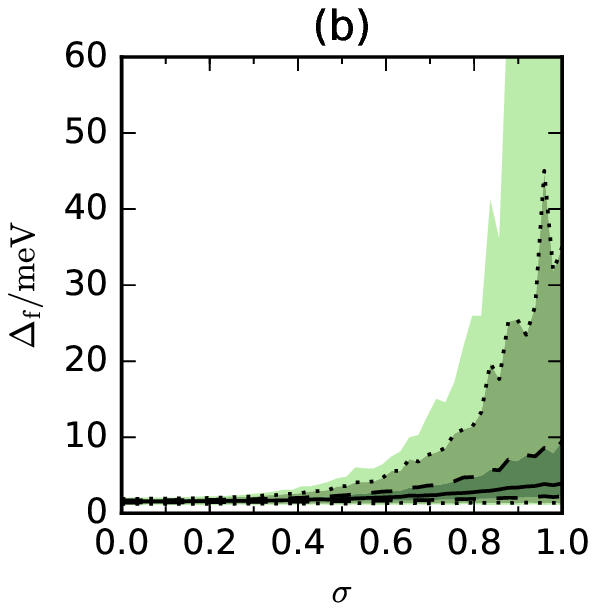} 
\includegraphics[width=0.3\textwidth]{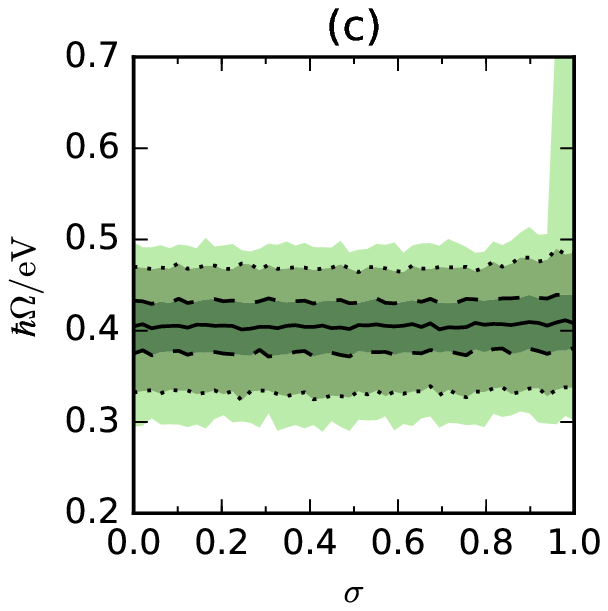}
\caption{Effects of the position randomness on the spectrum in a $4\times4$
dipole array ($16\,\mathrm{nm\times16\,\mathrm{nm}}$) for $\mu=2\,\mathrm{D}$,
$E=E_{z}=2.4\cdot10^{9}\,\mathrm{V\, m^{-1}}$. The dipoles are displaced
along each axis up to 1 nm. (a) center of the dipole band taken as
the mean of the second highest and the second lowest eigenvalue at
$\omega=\epsilon=2.6\,\mathrm{eV}$; (b) width of the dipole band
$\Delta_{f}$; (c) polariton splitting $\Omega$ from fitting the
relation (\ref{eq:Hopfield dispersion}). The lines correspond to
the quantiles 0.01, 0.05, 0.25, 0.5, 0.75, 0.95, 0.99, i.e. the areas
delimited by light, medium and dark shades delimit 98 \%, 90 \% and
50 \% of the values, respectively.\emph{\label{fig:pos}}}
\end{figure*}

\subsection*{Effects due to randomness}

As mentioned in the introduction, the QEs in the nanoplasmonic system
are usually distributed randomly near the metallic structures, having
also random directions. In order to capture the effects of the randomness,
we performed statistical simulations  with varying degree of randomness
in angular and positional configuration of the QEs. In both cases,
we start with a rectangular array of dipoles aligned in a single direction
(which corresponds to the direction of the field intensity). 

We choose several statistics calculated from the resulting spectra.
The width of the band can be described in multiple ways, one of them
is the difference $\Delta_{\mathrm{f}}$ between the second highest
and the second lowest eigenvalues at $\omega=\epsilon$. The average
$\epsilon_{\mathrm{m}}$ of these two eigenvalues characterises the
position of the band. For each sample, we perform a least-square fit
of the spectra onto the dispersion function (\ref{eq:Hopfield dispersion})
in order to obtain the parameters $\omega_{0}$ and $\Omega$, which
show the asymptotic behavior of the polariton branches and their splitting.\emph{
 }

As for the directions, the randomness is parametrized by the maximum
deviation polar angle $\Theta$. Each dipole is rotated from its aligned
direction by a random polar angle chosen uniformly between $0$ and
$\Theta$; the azimuth angle of the rotation is always chosen uniformly
from all directions. The resulting distributions of $\Delta_{\mathrm{f}}$
and $\Omega$ are illustrated in Fig. \ref{fig:angleRand} for a $4\times4$
square array with $2\,\mathrm{nm}$ space separation with two different
magnitudes of dipole moment, $2\,\mathrm{D}$ and $15\,\mathrm{D}$.
The dipole-field couplings were however kept in the same range of
$\pm0.26\,\mathrm{eV}$.  In all cases, the band center $\epsilon_{\mathrm{m}}$
was  equal to the QE natural frequency $\epsilon$ with a relative
error less than $2\cdot10^{-4}$. The fitted value of $\omega_{0}$
was equal to $\epsilon$ with 1\% accuracy (although always below
the prescribed $\epsilon$). \emph{}

The band width depends mainly on the magnitude of the dipole moment,
$\Delta_{\mathrm{f}}\propto\mu^{2}$. The directional randomness causes
variation in the band width, which in extreme cases can differ by
about a factor of two for different samples.

Next, we added some noise into the dipoles' positions. The initial
configuration was a $4\times4$ square array of randomly oriented
dipoles, with $a=2\,\mathrm{nm}$ distance between dipoles. However,
a random offset from the interval $(-\sigma a/2,\sigma a/2)$ was
then added to each cartesian coordinate of each dipole, where $\sigma$
is a randomness parameter. The resulting distributions of selected
statistics (for $\mu=2\mathrm{\, D}$) are shown in Fig. (\ref{fig:pos}).

The effect of the dipole configuration on the band position is again
negligible as it stays within a $1\,\mathrm{meV}$ range around the
original $\epsilon$ in 90 \% cases for the maximally random case.
However, the band width might increase substantially for some fraction
of samples in the maximally random case. This is caused by the fact
that the distance between two neighbouring dipoles can approach zero
and thus their mutual coupling $g_{ij}$ might become very large.
As will be discussed later, this situation is mostly unphysical because
of the nonzero size of the QEs.  The value of $\Omega$ is apparently
unaffected by the positions except for a very small fraction of cases,
which again correspond to unrealistically small distances between
the dipoles.

\subsection*{Scaling effects}

\begin{figure*}
\includegraphics[width=0.3\textwidth]{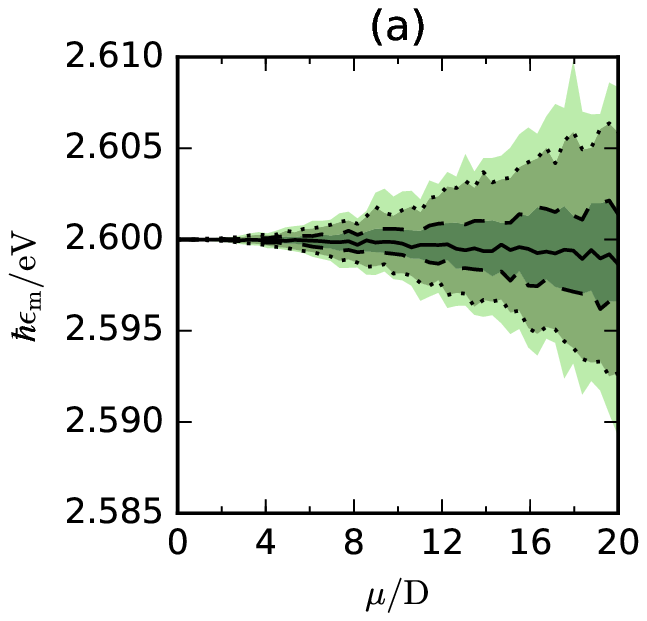}
\includegraphics[width=0.3\textwidth]{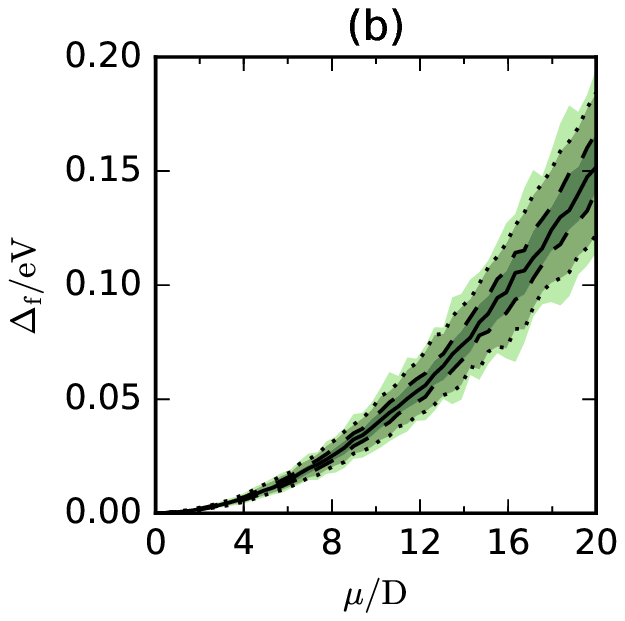}
\includegraphics[width=0.3\textwidth]{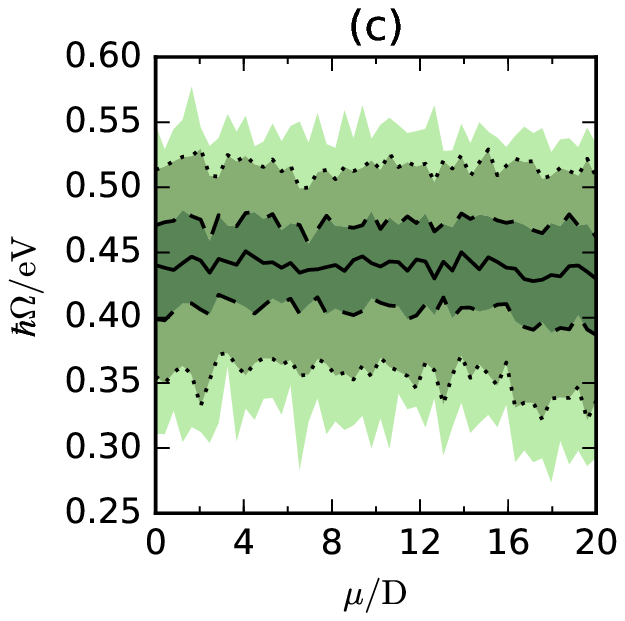}

\caption{Scaling of the observed statistics with the transition dipole moment
$\mu$\label{fig:Scaling-mu}, keeping the maximum field-dipole couplings
fixed at $V_{i}\le\mu E_{z}=0.1\,\mathrm{eV}$. Randomized are both
dipole orientations and positions. The lines correspond to the quantiles
0.01, 0.05, 0.25, 0.5, 0.75, 0.95, 0.99, i.e. the areas delimited
by light, medium and dark shades delimit 98 \%, 90 \% and 50 \% of
the values, respectively.}
\end{figure*}

In order to explore the effects of the direct dipole-dipole coupling
of the QEs, we scaled the transition dipole moment relevant for the
direct coupling while keeping the magnitude of $V_{i}$ coupling terms.
As stated before, increasing $\mu$ by the factor of $\alpha$ is
equivalent to reducing the intermolecular distance by the factor of
$\alpha^{-2/3}$. The results for the scenario with the $4\times4$
array of QEs with fixed positions at $2\,\mathrm{nm}$ interparticle
distance and fully random directions is showed in Fig. \ref{fig:Scaling-mu}.
As expected, the band width $\Delta_{\mathrm{f}}$ shows clear quadratic
dependence on $\mu$ (thus linear dependence on $g_{ij}$). The band
position remains well at the QE transition frequency $\epsilon$.
(Only the random fluctuations of the outermost energies of the band
scale linearly with the band width, which leads to the quadratic broadening
of the $\Delta_{\mathrm{f}}$ distribution with $\mu$.) The polariton
splitting $\Omega$ was found to be almost independent of dipole-dipole
interactions. Hence $\mu$ influences the polariton splitting only
via the $V_{i}\propto\vect{\mu}_{i}\cdot\vect E(\vect r_{i})$ terms.

\section{Conclusions\label{sec:Conclusions}}

A question naturally follows, whether the effects of the dipole-dipole
interaction described above can be probed experimentally in the nanoplasmonic
systems. The Hamiltonian (\ref{eq:Hamiltonian}), due to all the simplifications
made, describes a closed system without any coupling to a probing
field. We have shown, however, that the dark states are characterized
by very low expectation value of the observable $P$ defined by equation
(\ref{eq:The Observable-1}).

The experiments showing the strong coupling between the plasmonic
excitations and QEs are characterized by the observable polariton
splitting. While increasing the number of QEs and their couplings
to a plasmonic nanoparticle increases the polariton splitting, this
does not require any direct dipole-dipole interactions between the
QEs. We showed that if the dipole-dipole couplings between the QEs
are present, the previously degenerate QE transition energies split
into a broader band and some of the resulting states might radiate
much more intensively than others. However, this requires really significant
dipole-dipole coupling strengths. Larger dipole-dipole couplings can
be attained by increasing the QE concentration and/or transition dipole
moment. With a transition dipole moment $\mu\approx10\,\mathrm{D}$
and separations between the emitters of $\approx1\,\mathrm{nm}$,
there should be an observable band containing highly radiant states,
cf. Fig. \ref{fig:dark modes and observable}(d), but such values
might not be easy to attain. One of the most popular QEs used in active
nanoplasmonic systems is the rhodamine 6G (R6G) dye. The number density
of solid R6G is about $1.6\cdot10^{21}\mathrm{cm^{-3}}$ \citep{R6G_sciencelab},
corresponding to the intermolecular distance of $0.86\,\mathrm{nm}$.
The transition dipole moment of a separate R6G molecule is about 2
D \citep{penzkofer_orientation_1980}. These values correspond approximately
to the parameters of Fig. \ref{fig:dark modes and observable}(f),
where we might still expect some of the effects to be observable.
However, R6G is diluted or embedded into a polymer in the experiments.
The number density of $2.5\cdot10^{19}\,\mathrm{cm}^{-3}$ and typical
intermolecular separation of $3.5\,\mathrm{nm}$, corresponding to
the saturated water solution of R6G \citep{R6G_scbt}, thus provide
a more realistic estimate. For such parameters, the dipole-dipole
couplings are so small that no observable effects can be expected.

Based on these arguments, we can speculate that it would be challenging
but not impossible to observe effects of the direct quantum emitter
dipole-dipole couplings in the nanoplasmonic systems: it demands a
high value of the dipole moment concentration. Also in the paper of
Salomon \textit{et al.} \citep{salomon_strong_2012}, the new mode
appears in the absorption spectra for a high dipole moment value of
25 D and concentration of $10^{19}\,\mathrm{cm}^{-3}$. Note that
here we studied only the single excitation subspace. It is possible
that other important effects of the dipole-dipole interactions could
take place for higher excitation numbers.

\bibliographystyle{apsrev}
\bibliography{dipdip_cleaned,extra}

\begin{thebibliography}{20}
\expandafter\ifx\csname natexlab\endcsname\relax\def\natexlab#1{#1}\fi
\expandafter\ifx\csname bibnamefont\endcsname\relax
  \def\bibnamefont#1{#1}\fi
\expandafter\ifx\csname bibfnamefont\endcsname\relax
  \def\bibfnamefont#1{#1}\fi
\expandafter\ifx\csname citenamefont\endcsname\relax
  \def\citenamefont#1{#1}\fi
\expandafter\ifx\csname url\endcsname\relax
  \def\url#1{\texttt{#1}}\fi
\expandafter\ifx\csname urlprefix\endcsname\relax\def\urlprefix{URL }\fi
\providecommand{\bibinfo}[2]{#2}
\providecommand{\eprint}[2][]{\url{#2}}

\bibitem[{\citenamefont{Törmä and Barnes}(2015)}]{torma_strong_2015}
\bibinfo{author}{\bibfnamefont{P.}~\bibnamefont{Törmä}} \bibnamefont{and}
  \bibinfo{author}{\bibfnamefont{W.~L.} \bibnamefont{Barnes}},
  \bibinfo{journal}{Rep. Prog. Phys.} \textbf{\bibinfo{volume}{78}},
  \bibinfo{pages}{013901} (\bibinfo{year}{2015}).

\bibitem[{\citenamefont{Chikkaraddy et~al.}(2016)\citenamefont{Chikkaraddy,
  de~Nijs, Benz, Barrow, Scherman, Rosta, Demetriadou, Fox, Hess, and
  Baumberg}}]{chikkaraddy_single-molecule_2016}
\bibinfo{author}{\bibfnamefont{R.}~\bibnamefont{Chikkaraddy}},
  \bibinfo{author}{\bibfnamefont{B.}~\bibnamefont{de~Nijs}},
  \bibinfo{author}{\bibfnamefont{F.}~\bibnamefont{Benz}},
  \bibinfo{author}{\bibfnamefont{S.~J.} \bibnamefont{Barrow}},
  \bibinfo{author}{\bibfnamefont{O.~A.} \bibnamefont{Scherman}},
  \bibinfo{author}{\bibfnamefont{E.}~\bibnamefont{Rosta}},
  \bibinfo{author}{\bibfnamefont{A.}~\bibnamefont{Demetriadou}},
  \bibinfo{author}{\bibfnamefont{P.}~\bibnamefont{Fox}},
  \bibinfo{author}{\bibfnamefont{O.}~\bibnamefont{Hess}}, \bibnamefont{and}
  \bibinfo{author}{\bibfnamefont{J.~J.} \bibnamefont{Baumberg}},
  \bibinfo{journal}{Nature} \textbf{\bibinfo{volume}{535}},
  \bibinfo{pages}{127} (\bibinfo{year}{2016}).

\bibitem[{\citenamefont{Santhosh et~al.}(2016)\citenamefont{Santhosh, Bitton,
  Chuntonov, and Haran}}]{santhosh_vacuum_2016}
\bibinfo{author}{\bibfnamefont{K.}~\bibnamefont{Santhosh}},
  \bibinfo{author}{\bibfnamefont{O.}~\bibnamefont{Bitton}},
  \bibinfo{author}{\bibfnamefont{L.}~\bibnamefont{Chuntonov}},
  \bibnamefont{and} \bibinfo{author}{\bibfnamefont{G.}~\bibnamefont{Haran}},
  \bibinfo{journal}{Nat. Commun.} \textbf{\bibinfo{volume}{7}},
  \bibinfo{pages}{ncomms11823} (\bibinfo{year}{2016}).

\bibitem[{\citenamefont{Zengin et~al.}(2015)\citenamefont{Zengin, Wersäll,
  Nilsson, Antosiewicz, Käll, and Shegai}}]{zengin_realizing_2015}
\bibinfo{author}{\bibfnamefont{G.}~\bibnamefont{Zengin}},
  \bibinfo{author}{\bibfnamefont{M.}~\bibnamefont{Wersäll}},
  \bibinfo{author}{\bibfnamefont{S.}~\bibnamefont{Nilsson}},
  \bibinfo{author}{\bibfnamefont{T.~J.} \bibnamefont{Antosiewicz}},
  \bibinfo{author}{\bibfnamefont{M.}~\bibnamefont{Käll}}, \bibnamefont{and}
  \bibinfo{author}{\bibfnamefont{T.}~\bibnamefont{Shegai}},
  \bibinfo{journal}{Phys. Rev. Lett.} \textbf{\bibinfo{volume}{114}},
  \bibinfo{pages}{157401} (\bibinfo{year}{2015}).

\bibitem[{\citenamefont{Salomon et~al.}(2012)\citenamefont{Salomon, Gordon,
  Prior, Seideman, and Sukharev}}]{salomon_strong_2012}
\bibinfo{author}{\bibfnamefont{A.}~\bibnamefont{Salomon}},
  \bibinfo{author}{\bibfnamefont{R.~J.} \bibnamefont{Gordon}},
  \bibinfo{author}{\bibfnamefont{Y.}~\bibnamefont{Prior}},
  \bibinfo{author}{\bibfnamefont{T.}~\bibnamefont{Seideman}}, \bibnamefont{and}
  \bibinfo{author}{\bibfnamefont{M.}~\bibnamefont{Sukharev}},
  \bibinfo{journal}{Phys. Rev. Lett.} \textbf{\bibinfo{volume}{109}},
  \bibinfo{pages}{073002} (\bibinfo{year}{2012}).

\bibitem[{\citenamefont{Delga et~al.}(2014{\natexlab{a}})\citenamefont{Delga,
  Feist, Bravo-Abad, and Garcia-Vidal}}]{delga_quantum_2014}
\bibinfo{author}{\bibfnamefont{A.}~\bibnamefont{Delga}},
  \bibinfo{author}{\bibfnamefont{J.}~\bibnamefont{Feist}},
  \bibinfo{author}{\bibfnamefont{J.}~\bibnamefont{Bravo-Abad}},
  \bibnamefont{and}
  \bibinfo{author}{\bibfnamefont{F.}~\bibnamefont{Garcia-Vidal}},
  \bibinfo{journal}{Phys. Rev. Lett.} \textbf{\bibinfo{volume}{112}},
  \bibinfo{pages}{253601} (\bibinfo{year}{2014}{\natexlab{a}}).

\bibitem[{\citenamefont{Delga et~al.}(2014{\natexlab{b}})\citenamefont{Delga,
  Feist, Bravo-Abad, and Garcia-Vidal}}]{delga_theory_2014}
\bibinfo{author}{\bibfnamefont{A.}~\bibnamefont{Delga}},
  \bibinfo{author}{\bibfnamefont{J.}~\bibnamefont{Feist}},
  \bibinfo{author}{\bibfnamefont{J.}~\bibnamefont{Bravo-Abad}},
  \bibnamefont{and} \bibinfo{author}{\bibfnamefont{F.~J.}
  \bibnamefont{Garcia-Vidal}}, \bibinfo{journal}{J. Opt.}
  \textbf{\bibinfo{volume}{16}}, \bibinfo{pages}{114018}
  (\bibinfo{year}{2014}{\natexlab{b}}).

\bibitem[{\citenamefont{Dicke}(1954)}]{dicke_coherence_1954}
\bibinfo{author}{\bibfnamefont{R.~H.} \bibnamefont{Dicke}},
  \bibinfo{journal}{Phys. Rev.} \textbf{\bibinfo{volume}{93}},
  \bibinfo{pages}{99} (\bibinfo{year}{1954}).

\bibitem[{\citenamefont{Gaudin}(1976)}]{gaudin_diagonalisation_1976}
\bibinfo{author}{\bibfnamefont{M.}~\bibnamefont{Gaudin}}, \bibinfo{journal}{J.
  Phys. France} \textbf{\bibinfo{volume}{37}}, \bibinfo{pages}{1087}
  (\bibinfo{year}{1976}).

\bibitem[{\citenamefont{Pan et~al.}(2005)\citenamefont{Pan, Wang, Pan, Li, and
  Draayer}}]{pan_exact_2005}
\bibinfo{author}{\bibfnamefont{F.}~\bibnamefont{Pan}},
  \bibinfo{author}{\bibfnamefont{T.}~\bibnamefont{Wang}},
  \bibinfo{author}{\bibfnamefont{J.}~\bibnamefont{Pan}},
  \bibinfo{author}{\bibfnamefont{Y.-F.} \bibnamefont{Li}}, \bibnamefont{and}
  \bibinfo{author}{\bibfnamefont{J.~P.} \bibnamefont{Draayer}},
  \bibinfo{journal}{Phys. Lett. A} \textbf{\bibinfo{volume}{341}},
  \bibinfo{pages}{94} (\bibinfo{year}{2005}).

\bibitem[{\citenamefont{Sukharev and Nitzan}(2011)}]{sukharev_numerical_2011}
\bibinfo{author}{\bibfnamefont{M.}~\bibnamefont{Sukharev}} \bibnamefont{and}
  \bibinfo{author}{\bibfnamefont{A.}~\bibnamefont{Nitzan}},
  \bibinfo{journal}{Phys. Rev. A} \textbf{\bibinfo{volume}{84}},
  \bibinfo{pages}{043802} (\bibinfo{year}{2011}).

\bibitem[{\citenamefont{Wubs et~al.}(2004)\citenamefont{Wubs, Suttorp, and
  Lagendijk}}]{wubs_multiple-scattering_2004}
\bibinfo{author}{\bibfnamefont{M.}~\bibnamefont{Wubs}},
  \bibinfo{author}{\bibfnamefont{L.~G.} \bibnamefont{Suttorp}},
  \bibnamefont{and}
  \bibinfo{author}{\bibfnamefont{A.}~\bibnamefont{Lagendijk}},
  \bibinfo{journal}{Phys. Rev. A} \textbf{\bibinfo{volume}{70}},
  \bibinfo{pages}{053823} (\bibinfo{year}{2004}).

\bibitem[{\citenamefont{Fredkin and Mayergoyz}(2003)}]{fredkin_resonant_2003}
\bibinfo{author}{\bibfnamefont{D.}~\bibnamefont{Fredkin}} \bibnamefont{and}
  \bibinfo{author}{\bibfnamefont{I.}~\bibnamefont{Mayergoyz}},
  \bibinfo{journal}{Phys. Rev. Lett.} \textbf{\bibinfo{volume}{91}}
  (\bibinfo{year}{2003}).

\bibitem[{\citenamefont{Gruner and Welsch}(1996)}]{gruner_green-function_1996}
\bibinfo{author}{\bibfnamefont{T.}~\bibnamefont{Gruner}} \bibnamefont{and}
  \bibinfo{author}{\bibfnamefont{D.-G.} \bibnamefont{Welsch}},
  \bibinfo{journal}{Phys. Rev. A} \textbf{\bibinfo{volume}{53}},
  \bibinfo{pages}{1818} (\bibinfo{year}{1996}).

\bibitem[{\citenamefont{Ruppin}(1982)}]{ruppin_decay_1982}
\bibinfo{author}{\bibfnamefont{R.}~\bibnamefont{Ruppin}}, \bibinfo{journal}{The
  Journal of Chemical Physics} \textbf{\bibinfo{volume}{76}},
  \bibinfo{pages}{1681} (\bibinfo{year}{1982}).

\bibitem[{\citenamefont{Cohen-Tannoudji
  et~al.}(2000)\citenamefont{Cohen-Tannoudji, Dupont-Roc, and
  Grynberg}}]{cohen-tannoudji_processus_2000}
\bibinfo{author}{\bibfnamefont{C.}~\bibnamefont{Cohen-Tannoudji}},
  \bibinfo{author}{\bibfnamefont{J.}~\bibnamefont{Dupont-Roc}},
  \bibnamefont{and} \bibinfo{author}{\bibfnamefont{G.}~\bibnamefont{Grynberg}},
  \emph{\bibinfo{title}{Processus d'interaction entre photons et atomes}}
  (\bibinfo{publisher}{EDP Sciences}, \bibinfo{address}{Les Ulis France;
  Paris}, \bibinfo{year}{2000}), ISBN \bibinfo{isbn}{978-2-86883-358-7}.

\bibitem[{\citenamefont{Hopfield}(1958)}]{hopfield_theory_1958}
\bibinfo{author}{\bibfnamefont{J.~J.} \bibnamefont{Hopfield}},
  \bibinfo{journal}{Phys. Rev.} \textbf{\bibinfo{volume}{112}},
  \bibinfo{pages}{1555} (\bibinfo{year}{1958}).

\bibitem[{R6G({\natexlab{a}})}]{R6G_sciencelab}
\emph{\bibinfo{title}{Rhodamine {{6G}}}}, \bibinfo{howpublished}{{MSDS}
  {[online]}, {ScienceLab} (2013)}, \bibinfo{note}{accessed Aug 29, 2016},
  \urlprefix\url{http://www.sciencelab.com/msds.php?msdsId=9927579}.

\bibitem[{\citenamefont{Penzkofer and
  Wiedmann}(1980)}]{penzkofer_orientation_1980}
\bibinfo{author}{\bibfnamefont{A.}~\bibnamefont{Penzkofer}} \bibnamefont{and}
  \bibinfo{author}{\bibfnamefont{J.}~\bibnamefont{Wiedmann}},
  \bibinfo{journal}{Opt. Commun.} \textbf{\bibinfo{volume}{35}},
  \bibinfo{pages}{81} (\bibinfo{year}{1980}).

\bibitem[{R6G({\natexlab{b}})}]{R6G_scbt}
\emph{\bibinfo{title}{Rhodamine {{6G}}}}, \bibinfo{howpublished}{{Product
  database [online], {Santa Cruz Biotech}}}, \bibinfo{note}{accessed Aug 29,
  2016},
  \urlprefix\url{http://www.scbt.com/datasheet-280066-rhodamine-6g.html}.

\end{thebibliography}

\end{document}